# Spatial First-passage Statistics of Al/Si(111)-( $\sqrt{3} \times \sqrt{3}$ ) Step Fluctuations

B. R. Conrad, W. G. Cullen, D. B. Dougherty*, I. Lyubinetsky†, E. D. Williams

*University of Maryland, Department of Physics & MRSEC, College Park, Maryland 20742-4111*

## Abstract

Spatial step edge fluctuations on a multi-component surface of Al/Si(111)-( $\sqrt{3} \times \sqrt{3}$ ) were measured via scanning tunneling microscopy over a temperature range of 720K-1070K, for step lengths of L = 65-160 nm. Even though the time scale of fluctuations of steps on this surface varies by orders of magnitude over the indicated temperature ranges, measured first-passage spatial persistence and survival probabilities are temperature independent. The power law functional form for spatial persistence probabilities is confirmed and the symmetric spatial persistence exponent is measured to be θ = 0.498 ± 0.062 in agreement with the theoretical prediction θ = ½. The survival probability is found to scale directly with y/L, where y is the distance along the step edge. The form of the survival probabilities agree quantitatively with the theoretical prediction, which yields exponential decay in the limit of small y/L. The decay constant is found experimentally to be $y_s$/L= 0.076 ± 0.033 for y/L ≤ 0.2.

† Permanent address:  Pacific Northwest National Laboratory, EMSL, Richland, WA 99352

* Present address: Surface and Microanalysis Science Division, National Institute of Standards and Technology, 100 Bureau Dr. Mailstop 8372, Gaithersburg, MD 20899-8372

**Introduction**

The reasons for interest in nanoscale fluctuations stem from the drive to reduce the dimensions of electrical devices to a length scale that is comparable to defect fluctuation amplitudes.  In crystalline solids, the boundaries of layers of material, or monatomic step edges, are the dominant source/sink for atomic motion for the surfaces of crystalline solids[1-4].   In the regime where thermally activated atomic motion is allowed, the steps will change shape with time, or wander[5]. Traditionally, these step edge fluctuations have been examined using correlation function approaches. However additional information is available in the form of first-passage analyses[6-9], which may be pertinent to applications in self assembly and nanoscale device properties[10-13].

While first passage problems are most often posed in terms of temporal fluctuations, spatial wandering is also an applicable problem.  The distance that a fluctuation will persist along a step edge is particularly interesting as a measure of the stability of nanoscale structures[14-16]. Such information can be gained by examining spatial first passage statistics such as persistence and survival probabilities, P(y) and S(y) respectively.  Persistence probability P(y) is the probability that a fluctuating step edge does not return to its initial position over a given distance, y, measured parallel to the average step edge.  A closely related quantity, survival probability S(y) is the probability that a fluctuating step edge does not cross its average position over a given distance y. Formally, persistence and survival probabilities are defined as:

$$P_{ss}(y_\circ, y_\circ + y) \equiv Prob\{\ sign[x(y_\circ + y') - x(y_\circ)] = \text{constant}, \ \forall \ 0 \leq y' \leq y\} \quad (1)$$

$$S_{ss}(y_\circ, y_\circ + y) \equiv Prob\{\ sign[x(y_\circ + y') - \langle x \rangle] = \text{constant}, \ \forall \ 0 \leq y' \leq y\} \quad (2)$$

where x(y) is the displacement of the step, measured at a position y, from its average position. The brackets indicate an average over the length in question and it is assumed that steady state conditions exist and no growth is occurring[14].

Theoretical studies[12,14] have shown that persistence probabilities have the general form of a power law decay for the step displacement not to return to its starting position over a distance y,

$$P(y) \sim y^{-\theta} \qquad (3)$$

where θ is the persistence exponent [15, 17] characterizing the model universality class of the system. Fluctuations of step edges on Al/Si(111) display the time correlation function signature of a $t^{1/2}$ dependence at short times[18]. The most straightforward interpretation of this signature is that the fluctuations result from mass exchange randomly from all step edge positions with the neighboring terraces. Alternative explanations[19-21] based on diffusion limited kinetics are rendered less tenable in this case by the experimental cross correlation signature[22] and the observation of temporal persistence behavior consistent with z = 2 [23]. If the observations are due to random mass exchange, these dynamics fall within the Edwards-Wilkinson model, which can be described by the equation

$$\frac{\partial x(y,t)}{\partial t} = \Gamma \frac{\partial^2 x(y,t)}{\partial y^2} + \eta(y,t) \qquad (4)$$

where x is the step edge displacement, Γ is the mobility, and η is a noise term. We only consider [24]

$$\left\langle \eta(x,t)\eta(x',t') \right\rangle = \delta(x-x')\delta(t-t') \qquad (5)$$

which is uncorrelated Gaussian noise. It has been shown[16] that in the steady state configuration,

$$\theta = 3/2 - n, \quad 1/2 < n < 3/2$$
$$\theta = 0, \qquad n > 3/2 \tag{6}$$

where n = (z − d + 1)/2. For this (1+1) dimensional interface, d = 1 and the dynamical exponent of the Edwards-Wilkinson model is z = 2. Therefore, we expect the persistence exponent θ = ½ [15]. In comparison with the persistence probability, the survival probability is related to the autocorrelation function and decays roughly exponentially with decay constants related to the correlation length[14].

The spatial correlation function, i.e. the mean square displacement of a step edge as function of distance parallel to the edge, is defined as

$$G(y) \equiv \left\langle \left[ x(y - y_\circ) - x(y_\circ) \right]^2 \right\rangle_{y_\circ} \tag{7}$$

where the brackets indicate an average over an ensemble of initial step positions $y_0$. Using this definition, the spatial correlation function can be calculated directly from the measured step edge geometry, x(y). For small step edge distances, y smaller than the correlation length[25], the average of G(y) yields an initially linear behavior:

$$G(y) \sim \frac{kTy}{\tilde{\beta}} = \frac{b^2 y}{a} \tag{8}$$

where $\tilde{\beta}$ is the step edge stiffness, and $b^2/a$ is the step diffusivity. The experimental correlation function G(y) of every image was used to determine the linear region, over which persistence and survival probabilities were evaluated. As has been previously reported for the Al/Si(111) system[26], the step edge diffusivity for this data set follows a Boltzmann dependence on temperature, increasing from 0.45Å at 770 K to 1.00 Å at 1020 K.

**Experimental**

STM images were measured on Al/Si(111)-($\sqrt{3} \times \sqrt{3}$)R 30° surfaces at temperatures ranging from 720 K – 1070 K. Growth parameters were controlled to maintain the surface structure in the ($\sqrt{3} \times \sqrt{3}$)R 30° reconstruction induced by the deposition of Al onto the Si(111) surface[18, 26]. The experiments were conducted in a UHV chamber (base pressure $\sim 6 \times 10^{-11}$ Torr) equipped with a VT STM (Omicron), a rear-view LEED (Physical Electronics Industries), and a mass spectrometer (Pfeiffer Vacuum). The vicinal Si(111) Samples (As-doped, 10mΩ cm) were misoriented by 0.5° towards the $[11\bar{2}]$ direction. The Si surface was cleaned by several 5-s flashes at 1520 K with subsequent cooling at a slow rate ($\sim 20°/\text{min.}$) through the $(1 \times 1)$-to-$(7 \times 7)$ phase transition.

The Al/Si(111)-($\sqrt{3} \times \sqrt{3}$)R 30° reconstructed surface was prepared by evaporation of 0.25-0.33 ML of Al at a deposition rate of 0.5 ML/min on a Si substrate held at 1020 K[27, 28] and was monitored by LEED. The pressure during evaporation was below $3 \times 10^{-10}$ Torr and the Al flux was measured by a water cooled quartz microbalance (Leybold Inficon). The Si substrate was heated resistively with direct current while the temperature was measured via an infrared pyrometer. About 0.5 h of thermal stabilization was used before STM measurement at elevated temperatures.

The images chosen for this study were of two sizes, $(300nm)^2$ and $(500nm)^2$ with scan rates 3 μm/s and 15 μm/s. and pixel sizes 0.586 nm and 0.977 nm respectively. Where possible, only images that included enough monatomic steps to facilitate more than eight different step edge samplings were used for this analysis. Only single-layer steps were analyzed. For the analysis, the spatial STM images used must represent a 'snapshot' of the system, e.g. there should not be any significant edge dynamics occurring during the image acquisition [26]. At temperatures below 770 K, fluctuations are absent over time intervals of several minutes, while at 1020 K steps can fluctuate on the order of seconds[18]. Therefore, to obtain viable information above 870 K, samples were prepared at elevated temperatures and were then quenched at an initial cooling rate of over 200 K/s to room temperature in order to capture and preserve the step edge displacements[26].

A representative spatial image is presented in Fig. 1. Step displacements are defined by the x coordinate which is perpendicular to the direction of step edge propagation and a function of the y coordinate, which is parallel to the step edge[29]. The spatial deviations of each image's step edges x(y) are extracted after cropping the step edge of interest to eliminate any step regions that are marred by defects or pinning sites, and flattening the upper and lower terraces. Each constant y slice of the step edge image is fit to an analytic step-like hyperbolic tangent function and the inflection point of the function is extracted as the position of the step. A linear fit of the step positions is then subtracted from x(y) to account for a possible large scale wandering or rotation of the step edge. x(y) is then used to calculate correlation functions, autocorrelation functions, width distributions, persistence probabilities and survival probabilities. The indicated error bars are the standard deviations (one sigma) and are obtained from the deviations of repeated measurements.

The length of the step analyzed and the pixel size both are important as numerical simulations and theoretical calculations have shown that the persistence scales as $f(y/\delta y)$ as long as $y<L$[14, 30], and the survival scales as $f(y/L, \delta y/L)$ where L is the variable step length and δy is the image pixel size. Each step image used for this analysis was cropped from a larger original STM image, yielding a distribution of effective system sizes, L but the same value of the pixel size δy. For the entire data set, the range of values of δy/L was from 0.003 to 0.015. For the steps analyzed from any given image the smallest and largest values differed by no more than a factor of two.

**Results and Analysis**

Theoretical discussions implicitly assume that the equilibrium step displacements have a Gaussian distribution[15]

$$P(x) \propto \exp\left(-2\left[\frac{(x-x_o)}{w}\right]^2\right) \quad (9)$$

where $x_o$ is the maximum of the distribution and w is the width of the distribution. Using the measured values of x(y), the stationary single site height distributions were calculated and agree with a Gaussian functional form as shown Fig. 2 for data measured at 920 K. The fit yields a root-mean squared width of 2.37 ± 0.05 nm.

The persistence and survival probabilities were calculated as described above over the temperature range 720 K-1070 K. Examples of a linear plot of persistence and survival probabilities versus distance parallel to the step edge, y, are shown in Fig. 3. The same persistence curve with a power law fit using logarithmic scales is shown in the inset to more clearly illustrate the date. Deviations to the power law fit occur outside the linear region of the correlation function and therefore do not appreciably effect persistence exponent measurements. The deviations themselves stem from limited statistics at large y as well as possible effects of finite measurement size issues, as discussed below. The average of the persistence curves for all the steps in one image is fit to Eq. 3 to extract the persistence exponent θ. Fig. 4 is a linear plot of the persistence exponent values versus the temperature. No systematic dependence on temperature is observed, and a weighted linear fit of the persistence exponent versus temperature produces a slope close to zero, $-7.7 \times 10^{-5} \pm 2.7 \times 10^{-5}$ K$^{-1}$. An analysis of the averaged persistence probabilities over all the temperatures results in a persistence exponent of θ = 0.498 ± 0.062.

The survival curves are found empirically to follow an exponential decay at small distances:

$$S(y) \sim \exp(^{-y}\!/\!_{y_s}) . \qquad (10)$$

The measured survival length constant showed a great deal of scatter, with no apparent correlation with changes in temperature. A weighted linear fit of the survival length constant versus temperature produces a slope of $1.2 \times 10^{-3} \pm 6.6 \times 10^{-3}\, nm * K^{-1}$, and an average value of 9.2 ± 5.7 nm. More physical analysis requires correcting for the fact that each measurement was carried out for a step segment of a different length. It is known that survival probability can be described by a scaling function[15]

$$S(y, L, \delta y) = f(y/L, \delta y/L) \qquad (11)$$

where L is the size of the system and δy is the pixel size of the image. Therefore the survival curves for the individual steps in each image were calculated as a function of *y/L*, and then fit as *S(y/L) ~ exp[(y/L/(y_s/L)]*. The individual length constants *y_s/L* for each of the steps in one image were then averaged to give the average scaled survival length constant for the image. The scaled survival length constants *y_s/L* are plotted versus temperature in Fig. 5. The average scaled survival length constant is found to be 0.076 ± 0.033 and a weighted linear fit of the scaled survival length versus temperature produces a slope of $-3.5 \times 10^{-5} \pm 7.0 \times 10^{-5} \, K^{-1}$, e.g. any true temperature variation must be smaller in magnitude than the experimental uncertainty in the data.

To illustrate the effects of step-length scaling, data measured for individual steps with different pixel size and a wide range of step lengths are shown in Fig. 6a. The collapse of the scaled survival probability curves with scaling as $y/L$ is shown in Fig. 6b. For large distances y, the survival probability statistics significantly decrease and variations between measurements and deviations from the theory are observed. By analogy with the effects of finite measurement times[31], such deviations of the survival probability may be expected for large distances y due to the finite sample size. No systematic effect of the pixel ratio on the linear region is observed in Fig. 6. This is confirmed by evaluating of the variation of the scaled decay length $y_s/L$ with pixel ratio, $\delta y/L$, for all the steps analyzed. The result showed no systematic dependence over the measurement range of $0.003 < \delta y/L < 0.015$.

**Discussion and Conclusions**

In general, the spatial data obtained in this study is noisier than in previous temporal studies[18, 22, 23, 26]. Nevertheless, the measured persistence exponent value of $\theta$ = $0.498 \pm 0.062$ is clearly in agreement with the theoretical value of ½. As can be seen in Figure 4, there is no apparent temperature dependence of the exponent. This indicates that there is no change in the value of z in Eq. 6, and thus no change in the underlying mechanism of the step motion over the temperature range as observed previously[22]. This lack of temperature dependence is consistent with the previous determination of the temporal persistence exponent for this system[23].

The survival probability curves have been shown to scale with system size as expected, and to follow an exponential decay at small distances. Full theoretical predictions are available for the spatial survival to longer distance scales, which can be written as an expansion[15]:

$$S(u) = 1 - \frac{4\sqrt{3u}}{\pi} + \frac{8}{\pi\sqrt{3}}u^{3/2} + \frac{4\sqrt{3}}{\pi}u^{5/2} - \frac{32\sqrt{3}a(1-a)}{\pi}u^{7/2} \quad (12)$$

where the parameter a = ½ and the scaled length parameter is y/L = u. This curve is shown as the solid line plotted in Fig. 6, and reproduces the rapid fall-off of the survival probability at larger distances. Consistent with the experimental observation, the functional form is indistinguishable from an exponential for y/L <~ 0.2. A fit of the theoretical curve by an exponential over similar length scales provides a scaled survival length constant of $y_s/L = 0.122$, somewhat larger than the average measured value of $0.076 \pm 0.033$. This empirical survival length constant is a useful experimental rule of thumb. This constant is independent of sample dependent system length, and provides ratio of the characteristic fluctuation length scales to the system size[15]. Furthermore, an analysis of all the data, illustrated for a subset of the data in Fig. 6, show that the scaled decay length for the linear exponential region of the fit is robust with respect to changes of a factor of 5 in the pixel size.

In summary, spatial first-passage statistics have been used to analyze step fluctuations on Al/Si(111). The temperature dependent study on a model metal-semiconductor surface was carried out on a variable-temperature STM. The quantitative examination of step fluctuation dynamics was based on analysis of both traditional spatial correlation functions and the statistically based persistence and survival. The stationary displacement distribution of the step deviations is confirmed to have a Gaussian functional form as predicted. The extracted mean squared width provides valuable information concerning the average step edge displacement.

However, when this information is combined with the predictive nature of persistence and survival studies the experimentally meaningful length scales are easily extracted. The spatial persistence exponent is measured to be $0.498 \pm 0.062$ in agreement with the theoretical prediction of ½ for the Edwards-Wilkins model.  This is further confirmation that the step fluctuations in this system are governed by random exchange of mass with the terraces over the entire temperature range of observation. An effective exponential form for the survival probabilities is found, with a scaled survival length constant value of $0.076 \pm 0.033$. The survival probability is observed to scale directly with y/L, where y is the step edge position and L is the step length, and the overall shape of the curve agrees well with theoretical prediction[15].  Both the extracted persistence exponent and survival length constant are observed to be temperature independent over the range 720K-1070K, where the underlying mass transport rates in this system change by three orders of magnitude[18]. This can be traced back to the finite sample size of the measurements. Since the step length is playing the role of the correlation length, the temperature independence of the persistence exponent and the survival length constant is expected.

**Acknowledgement**

This work has been supported by the University of Maryland NSF-MRSEC under grant DMR 05-20471.  The NSF-MRSEC SEF was used in obtaining the data presented.

Useful discussions with Professor Chandan Das Gupta and Blake Riddick are also gratefully acknowledged.

**Figures**

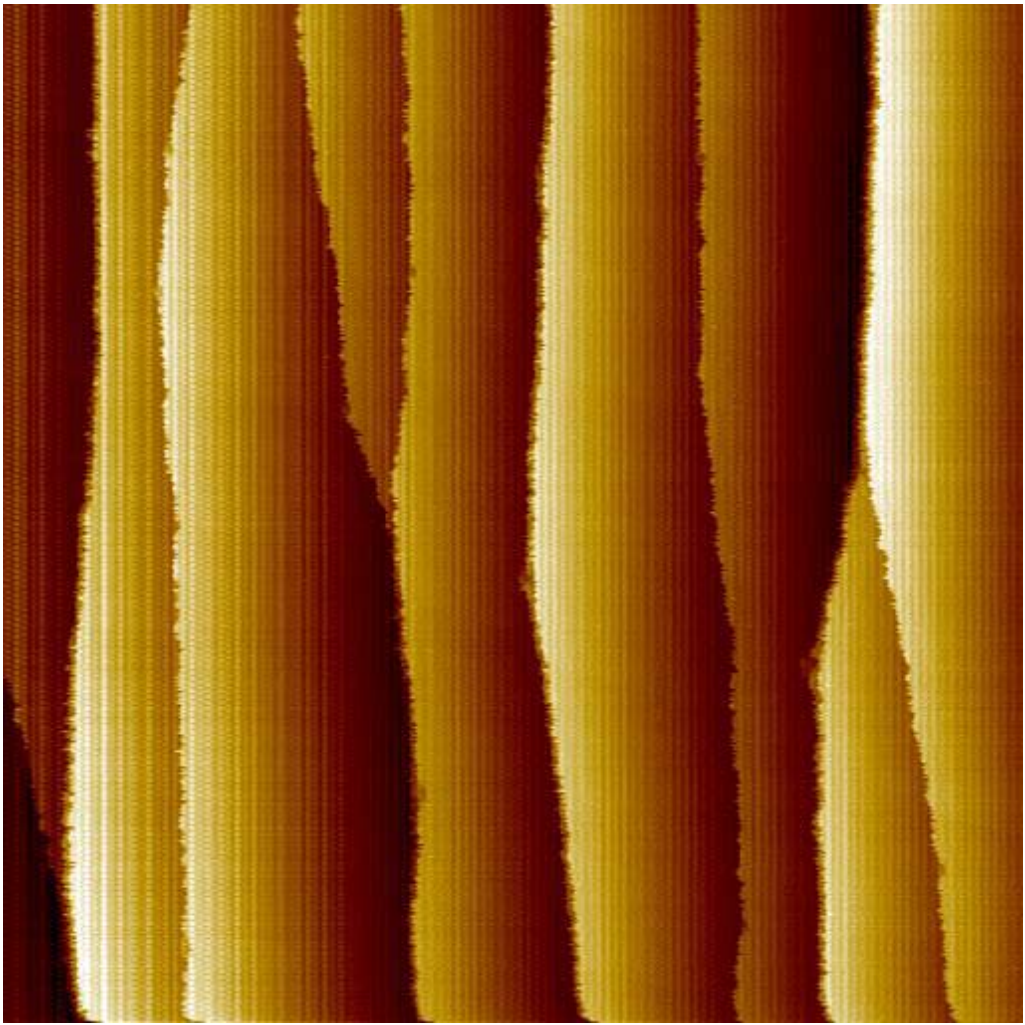

Figure 1. (color online) A $\left(500\ nm\right)^2$ STM image of Al/Si(111) with pixel size of 0.977

nm, measured at 970 K.  The single step heights are 3.1Å.

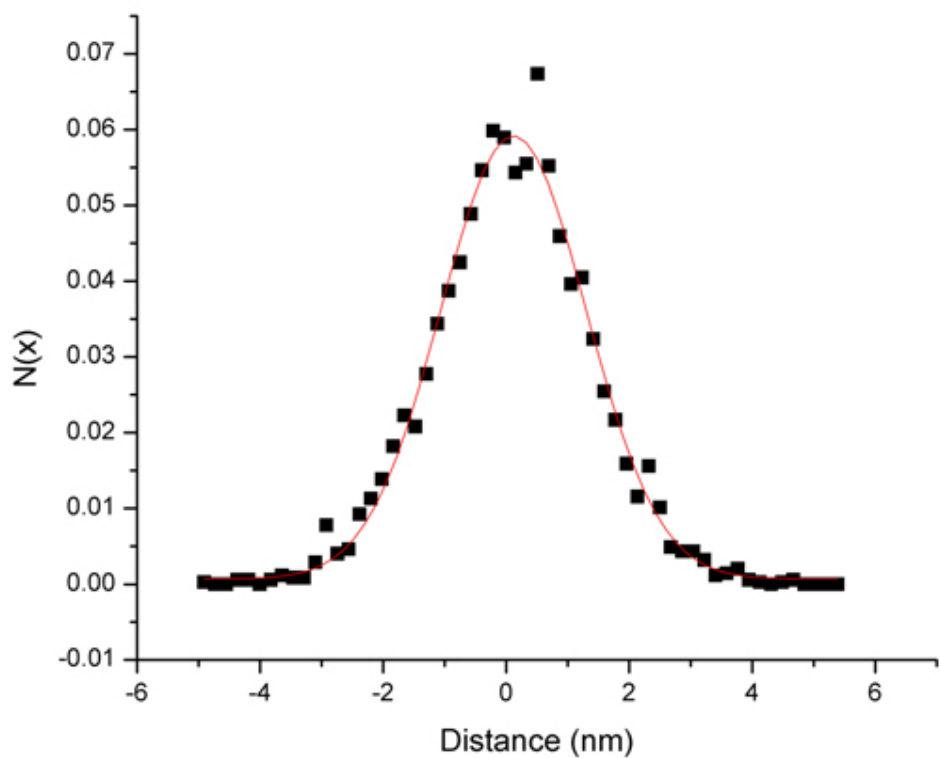

Figure 2.  Stationary single site height distribution for all data taken at 920 K. The fitting

parameters are $x_o = 0.118 \pm 0.020$ nm and $w = 2.37 \pm 0.05$ nm.

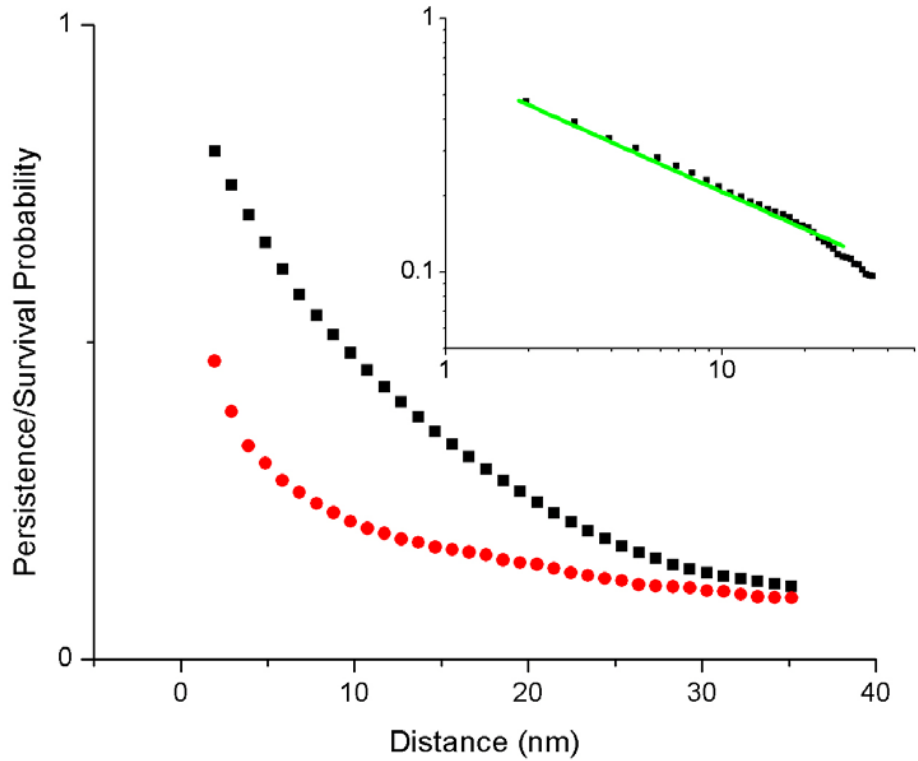

FIG. 3.  (color online)  Representative persistence and survival probability data. The data were taken at 970 K, from an STM image with pixel size of 0.977 nm. The persistence and survival curves are represented by squares and circles respectively. The inset is the same persistence curve using logarithmic scales. The solid green line is a power law fit to the data over the linear region of the spatial correlation function with the persistence exponent θ = 0.59.  Error bars are 1-sigma values of measurements on seven to ten step segments each measured from the steps in a single STM image.  The true standard deviation would be obtained in the limit of a large number of such measurements, and here is estimated by a sampling of several such images.

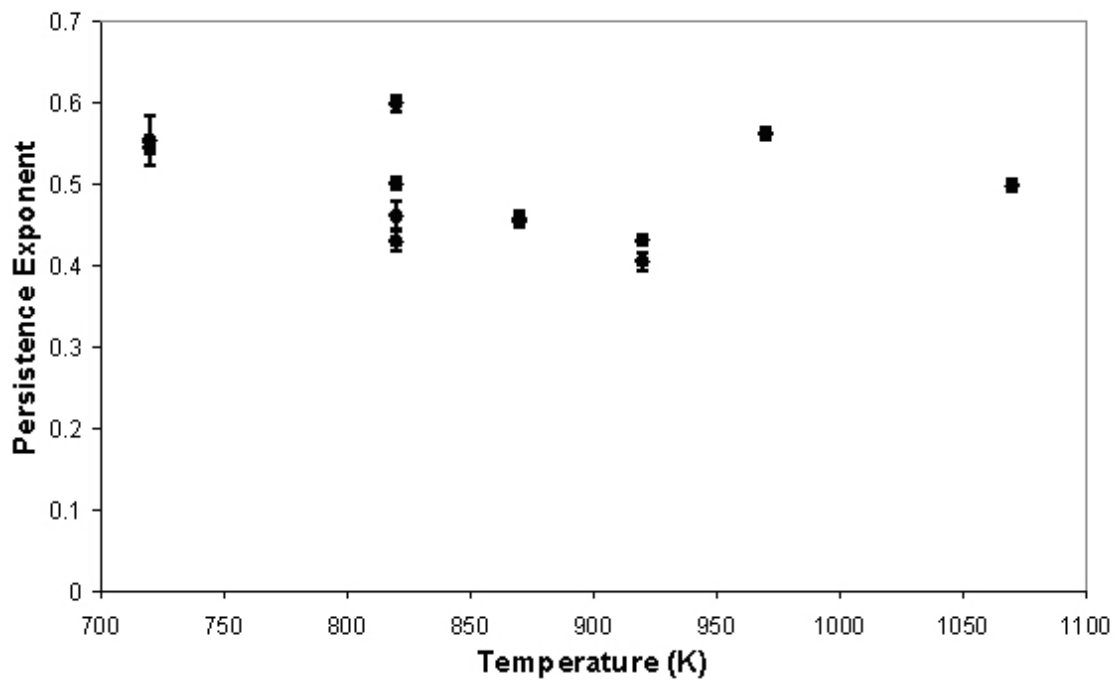

FIG. 4. Persistence exponents vs. temperature. Error bars are 1-sigma values of measurements on seven to ten step segments each measured from the steps in a single STM image. The true standard deviation would be obtained in the limit of a large number of such measurements, and here is estimated by a sampling of several such images.

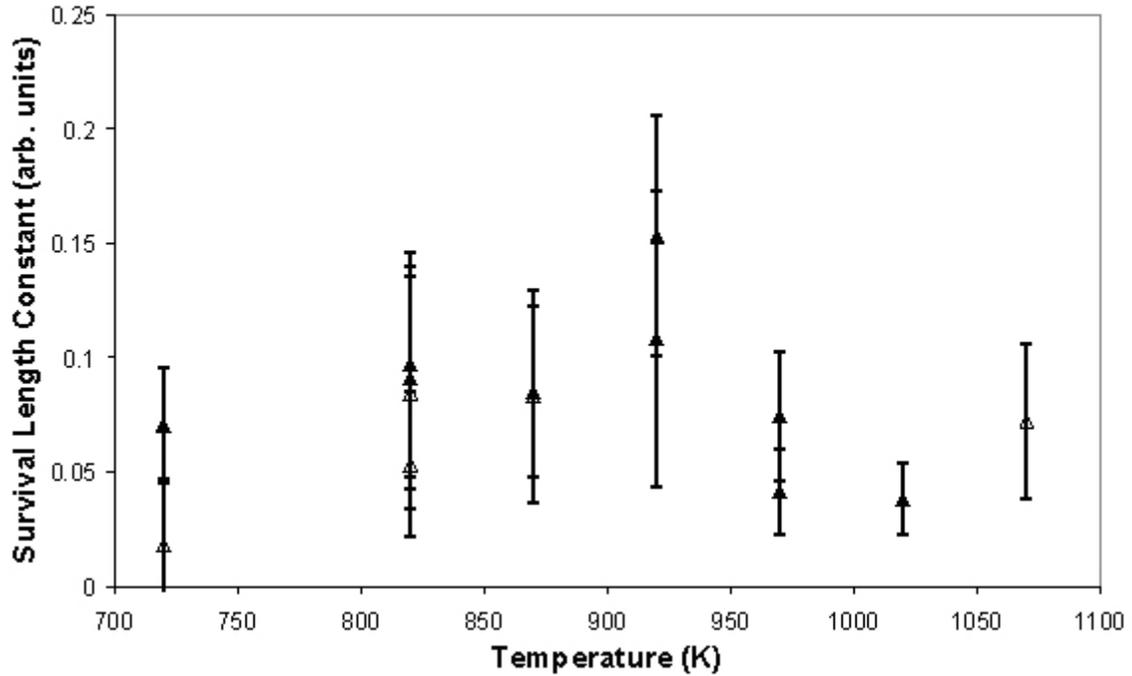

Figure 5. Scaled survival decay length vs. temperature. Error bars are 1-sigma values of measurements on eight to ten step segments. Open and solid triangles are from

$(300\ nm)^2$ images and $(500\ nm)^2$ images respectively. The lengths of the steps analyzed were

37.5 - 134 nm (720 K), 73 – 237 nm (820 K), 56 – 111 nm (870 K), 97 – 277 nm (920 K),

87 – 139 nm (970 K), 63 – 124 nm (1020 K), 118 – 194 nm (1070 K).

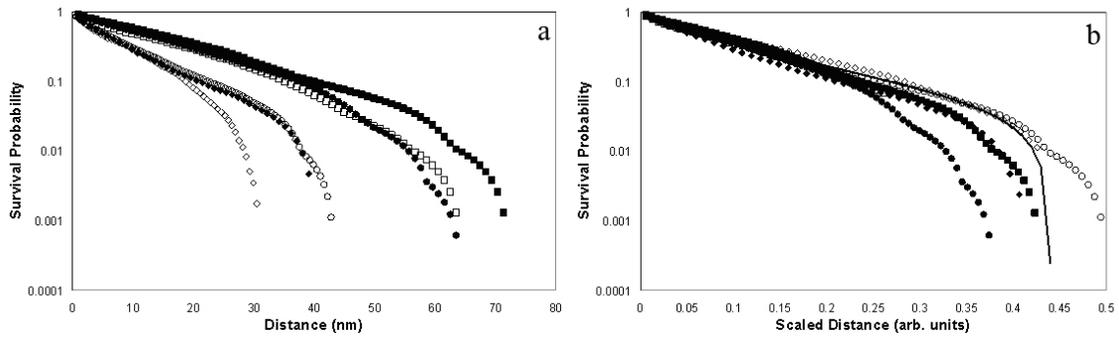

Figure 6. Survival probabilities determined from single steps chosen to display measurements at both pixel sizes and a wide range of step lengths. Solid diamonds, squares, and circles are from $(500\ nm)^2$ images and have system lengths of L = 98.9 nm, 170 nm, and 162 nm respectively. Open diamonds, squares, and circles are from $(300\ nm)^2$ images and have system lengths of L = 65.8 nm, 154 nm, and 87 nm respectively. a) Survival probabilities vs. distance, y, parallel to the step edge. b) Survival probabilities vs. scaled distance, y/l. The solid line is the theoretical prediction of Eq. 12 [15].